\documentstyle[prb,aps,epsf,multicol]{revtex}

\newcommand{\bleq}{\ifpreprintsty
                   \else
                   \end{multicols}\vspace*{-3.5ex}{\tiny
                   \noindent\begin{tabular}[t]{c|}
                   \parbox{0.493\hsize}{~} \\ \hline \end{tabular}}
                   \fi}
\newcommand{\eleq}{\ifpreprintsty
                   \else
                   {\tiny\hspace*{\fill}\begin{tabular}[t]{|c}\hline
                    \parbox{0.49\hsize}{~} \\
                    \end{tabular}}\vspace*{-2.5ex}\begin{multicols}{2}
                    \fi}
\newcommand{\bcols}{\ifpreprintsty\else\begin{multicols}{2}\fi}
\newcommand{\ecols}{\ifpreprintsty\else\end{multicols}\fi}

\begin{document}
\draft

\title{Coulomb blockade threshold in inhomogeneous one-dimensional arrays of 
tunnel junctions}
\author{J. A. Melsen} 
\address{Instituut-Lorentz, University of Leiden\\
P.O. Box 9506, 2300 RA Leiden, The Netherlands}
\author{Ulrik Hanke, H.-O. M\"uller, and K.-A. Chao}
\address{Department of Physics and Mathematics, Norwegian
Institute of Technology\\ University of Trondheim, 7034 Trondheim, Norway}

\maketitle

\begin{abstract}
A general expression is given for the change in free energy when
a charge tunnels through a junction in a one-dimensional array of
$N$ metallic islands with arbitrary capacitances and arbitrary 
background charges. This is used to obtain expressions for the (average)
threshold voltage of the Coulomb blockade for a few characteristic geometries.
We find that including random background charges has a large effect on the 
$N$-dependence of the threshold voltage:
In an array with identical junction capacitances $C$ and gate 
capacitances $C_g$, the threshold voltage, averaged over the background 
charge, is proportional to $N^a$, 
where $a$ crosses over from $\case{1}{2}$ to 1 when $N$ becomes larger than
$2.5\sqrt{C/C_g}$.
\bigskip
\end{abstract}


\bcols 

\section{Introduction}
Since the pioneering work by Gorter in 1951,\cite{Gorter} single charge 
tunneling effects have been extensively studied in various kinds of 
geometries.\cite{review} Research on single electronics has led to potential
applications in e.g. current standards,\cite{Geerligs,Pothier}
ultradense integrated digital electronics,\cite{Korotkov}
thermometry,\cite{Pekola,Hirvi} and room-temperature memory.\cite{Yano} In many 
of these applications, tunneling occurs through a large number of junctions in 
series. Most theoretical work has assumed homogeneous 
arrays.\cite{Bakhvalov,BenJacob,Averin93,Hu,Kang} The problem is that the 
number of available states at a finite current rapidly increases with the 
circuit size, so that one either restricts the analysis to homogeneous arrays 
or adopts a numerical approach.\cite{Fonseca} Using modern techniques, it is 
possible to fabricate arrays of metallic islands separated by tunnel junctions 
with almost uniform capacitances. It is however very difficult to avoid 
non-uniform background charges on the islands. This is relevant, since the 
charging energy is very sensitive to the background charge.

The aim of this article is to provide results for {\em inhomogeneous} 
one-dimensional arrays of metallic islands. The inhomogeneity can be both in
the junction capacitances and in the background charges on the islands in the
array. In particular, we study the threshold voltage for charge transport.
The results obtained are exact within the classical (orthodox) model of single
electron tunneling,\cite{Averin} which is accurate when quantum size effects 
and macroscopic quantum tunneling effects may be ignored.

Using a general expression for the inverse capacitance matrix, we calculate
in Section \ref{sec:two} the change in the free energy of an $N$-junction
array due to an arbitrary tunneling event. In Section \ref{sec:three}, we focus
on the threshold voltage for transport $V_t$, which is an observable quantity.
We find that inhomogeneity of the junction capacitances $C$ has a small effect
on the threshold voltage in large arrays: The expectation value as $N\to\infty$
for the threshold voltage of an array without gate coupling (gate capacitance
$C_g=0$ for each junction) and without 
background charges is $\langle V_t\rangle = \case{1}{2}Ne\langle C^{-1}\rangle$,
with $\langle C^{-1}\rangle$ being typically not much different from 
$1/\langle C\rangle$. However, as we show in Section \ref{sec:four}, a
random variation in background charges may change the threshold voltage 
considerably: In a short array with weak gate coupling ($N^2C_g/6.25 C < 1$) 
and random 
charges on all $N$ islands, we find $\langle V_t\rangle \propto \sqrt{N}$. In a
long array with strong gate coupling ($N^2C_g/6.25 C\gg 1$, but still 
$C_g \ll C$),
we find $\langle V_t\rangle \propto N$. We compare our results with 
experiments.\cite{Delsing}
 
\section{Free energy}
\label{sec:two}

\begin{figure}[htb]
\epsfxsize=0.95\hsize
\vspace*{-3ex}\epsffile{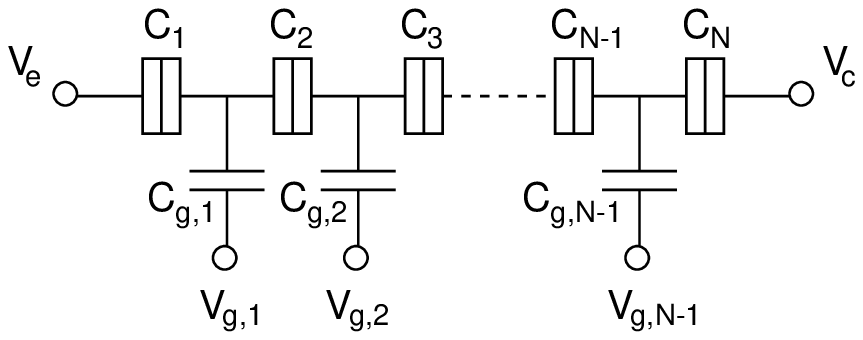}\vspace*{0ex}
\refstepcounter{figure}
\label{fig:array}
{\small FIG. \ref{fig:array}.
Schematic diagram of a one-dimensional array of 
$N$ tunnel junctions. Island $i$ is coupled to island $i+1$ by a tunnel 
barrier with capacitance $C_{i+1}$, and to a gate electrode by an insulating 
barrier with capacitance $C_{g,i}$. The capacitance $C_1$ ($C_N$)
denotes the coupling of the first (last) island to the emitter (collector)
electrode.}
\end{figure}%
\noindent
The system under consideration is shown schematically in Fig.\ \ref{fig:array}.
Within the orthodox model, the state of the system is described by the 
numbers $n_i$ of electrons on the $i$-th island, which we combine in a
vector: $\vec{n}\equiv(n_1,n_2,\ldots,n_{N-1})$. The tunneling rate, 
$\Gamma_k(\vec{n})$, corresponding to a single electron tunneling from 
island $k-1$ to island $k$ is given by\cite{review}
\begin{equation}
\Gamma_k(\vec{n})=\frac{\Delta G_k(\vec{n})}{e^2R_k[1-\exp(-\Delta G_k(\vec{n})/
k_{\rm B} T)]}.
\end{equation}
Here $R_k$ is the resistance of the $k$-th tunnel junction and 
$\Delta G_k(\vec{n})$ is defined as the difference
in free energy of the final and initial states. The free energy comprises the 
electrostatic energies of the charged capacitors in the system, as well as the 
potential energies of all electrodes:\cite{Bakhvalov}
\begin{eqnarray}
G(\vec{n}) &  = & \frac{1}{2}\sum_{i=1}^{N-1}C_{g,i}(\phi_i-V_{g,i})^2+
       \frac{1}{2}\sum_{i=1}^N C_i(\phi_i-\phi_{i-1})^2\nonumber\\
&&-V_e Q_e-V_c Q_c- \sum_{i=1}^{N-1}V_{g,i}Q_{g,i}.
\label{eq:Gibbs}
\end{eqnarray}
We denote by $\phi_i$ the electrochemical potential of island $i$
($\phi_0\equiv V_e$ and $\phi_N\equiv V_c$), and by $Q_e$, $Q_c$, and 
$Q_{g,i}$ the charges on the emitter, collector, and gates, respectively:
\begin{mathletters}
\begin{eqnarray}
Q_e & = & C_1(V_e-\phi_1)+e n_e,\\
Q_c & = & C_N(V_c-\phi_{N-1})+e n_c,\\
Q_{q,i} & =  & C_{g,i}(V_{g,i}-\phi_i).
\end{eqnarray}
\end{mathletters}%
Here $n_e$ ($n_c$) is the number of electrons that has tunneled from the 
emitter (collector) electrode through the first (last) capacitor. 

The difficulty in determining the energy difference $\Delta G_k(\vec{n})$ 
lies in the determination of the electrochemical potentials $\vec{\phi}\equiv
(\phi_1,\phi_2,\ldots,\phi_{N-1})$. They follow from the condition that 
the total capacitive charge on each island $i$ equals $en_i$ plus a background
charge $Q_{0,i}$:
\begin{eqnarray}
&&C_{g,i}(\phi_i-V_{g,i})+C_i(\phi_i-\phi_{i-1})+C_{i+1}(\phi_i-\phi_{i+1})
\nonumber\\
&& = en_i+Q_{0,i},\quad i=1,2,\ldots,N-1.
\label{eq:chargecons}
\end{eqnarray}
The background charge $Q_{0,i}\in (-e/2,e/2)$ is due to incompletely screened 
charges in the environment of the island.  Eq.\ (\ref{eq:chargecons}) can be
written in matrix form as $\mbox{\boldmath $C$}\vec\phi  =  \vec{Q'}$, with
\begin{mathletters}
\begin{eqnarray}
&& \mbox{\boldmath $C$}_{ij} = \delta_{i,j}(C_i+C_{i+1}+C_{g,i})-
                                \delta_{i+1,j}C_j+\delta_{i,j+1}C_i,\\
&&Q'_i = e n_i+Q_{0,i}+C_{g,i}V_{g,i}+
           \delta_{i,1}C_1V_e+\delta_{i,N-1}C_NV_c.\nonumber\\
&& 
\end{eqnarray}
\end{mathletters}%
The capacitance matrix $\mbox{\boldmath $C$}$ can be inverted exactly. The 
elements $R_{i,j}$ of the inverse capacitance matrix 
$\mbox{\boldmath $R$}=\mbox{\boldmath $C$}^{-1}$ are given by
\begin{eqnarray}
R_{i,j} & = & C_{i+1}C_{i+2}\cdots C_j D_{i-1} \tilde{D}_{j+1} D_{N-1}^{-1},
\quad i \le j,\nonumber\\
R_{j,i} & = & R_{i,j}.
\label{eq:Rel}
\end{eqnarray}
Here we have introduced the subdeterminants $D_i$ ($\tilde{D}_{N-i}$) of the 
upper left (lower right) capacitance submatrix of dimension $i$. These 
can be found recursively from
\begin{mathletters}
\label{eq:recursiveDs}
\begin{eqnarray}
&& D_i = (C_i+C_{i+1}+C_{g,i})D_{i-1}-C_i^2D_{i-2},\label{eq:recursiveD}\\
&&\tilde{D}_i= (C_i+C_{i+1}+C_{g,i})\tilde{D}_{i+1}-C_{i+1}^2\tilde{D}_{i+2},
\label{eq:recursivetildeD}\\
&&D_0 \equiv  \tilde{D}_N \equiv 1.
\end{eqnarray}
\end{mathletters}%
\noindent
For a homogeneous array with identical capacitances,
$C_1=C_2=\ldots=C_N$ and $C_{g,1}=C_{g,2}=\ldots=C_{g,N-1}$, we 
recover the inverse capacitance matrix of Ref.\ \onlinecite{Hu}.

We now derive a general expression for the difference in free energy 
$\Delta G_k(\vec{n})$ when an electron tunnels from island $k-1$ to island $k$.
Applying Eq.\ (\ref{eq:recursiveDs}) and making use of the orthogonality 
relation 
\begin{equation}
(C_i+C_{i+1}+C_{g,i})R_{i,j}=C_iR_{i-1,j}+ C_{i+1}R_{i+1,j}+\delta_{i,j},
\label{eq:orthogonality}
\end{equation}
we find that $\Delta G_k(\vec{n})$ takes the form
\bleq
\begin{mathletters}
\label{eq:dG}\\
\begin{eqnarray}
\Delta G_k(\vec{n})&=&-\frac{e^2}{2}(R_{k-1,k-1}+R_{k,k}-R_{k-1,k}-R_{k,k-1})+
e\sum_{i=1}^{N-1}Q_i(R_{i,k-1}-R_{i,k})\nonumber\\
& & \mbox{}+e(V_e-V_{g,1})A_{1,k}+ e\sum_{i=2}^{N-1}(V_{g,i-1}-V_{g,i})A_{i,k}
+e(V_{g,N-1}-V_c)A_{N,k},\\
A_{i,k} & = & C_i(R_{i-1,k}+R_{i,k-1}-R_{i-1,k-1}-R_{i,k})+\delta_{i,k}.
\end{eqnarray}
\end{mathletters}%
\eleq
\noindent
Here, $R_{i,N}=R_{0,i}=0$ is implied.

Although we are now able to construct all relevant transition rates from 
expressions (\ref{eq:Rel}) and (\ref{eq:dG}), the analytic evaluation of 
the current-voltage characteristic at arbitrary voltage remains a technically 
involved problem.  The threshold voltage, however, is determined by 
a single transition rate and is therefore easier to evaluate.
In the next two sections, we apply our results to this quantity
for several characteristic geometries.

\section{Threshold voltage}
\label{sec:three}

Electron transport through a one-dimensional array is realized by a 
sequence of tunneling events through all junctions between the emitter and the 
collector (we refer to this as a tunneling sequence). At zero temperature, 
a specific tunneling sequence contributes to the conductance if the free energy 
difference of each tunneling event in the sequence is positive.
The threshold voltage $V_t$ of the Coulomb blockade is the smallest voltage
at which a current can flow through the array at zero temperature.
When $|V_e-V_c| < |V_t|$, there exists no conductive tunneling sequence.
We first consider the simple case where the system is not gated ($C_{g,i}=0$
for all $i$), and then discuss the turnstile configuration,
i.e. an array which is coupled to a gate electrode via a single island: 
$C_{g,i}=C_g\delta_{i,n}$. 

\subsection{No gate coupling}
In the absence of gate coupling, the determinants $D$ and $\tilde{D}$,
following from Eq.\ (\ref{eq:recursiveDs}), have a simple form. For 
convenience, we introduce the notation
\begin{equation}
\label{eq:invsum}
S_k^l\equiv\sum_{i=k+1}^l\frac{1}{C_i},\quad S^l\equiv S_0^l,
\quad S_k\equiv S_k^N.
\end{equation}
In terms of these quantities,
\begin{mathletters}
\begin{eqnarray}
&&D_k=C_1C_2\cdots C_{k+1} S^{k+1},\\
&&\tilde{D}_k=C_kC_{k+1}\cdots 
C_N S_{k-1},\\
&&R_{i,j}=S^i S_j / S^N, i\leq j.
\end{eqnarray}
\end{mathletters}%
We further define $\vec{q}\equiv \vec{n}+\vec{q}_0$, 
$\vec{q}_0\equiv e^{-1}(Q_{0,1},Q_{0,2},\ldots,Q_{0,N-1})$.
From the condition $\Delta G_k(\vec{q})=0$, we determine the threshold voltage
$V_{t,k}(\vec{q})$ for tunneling through capacitance $C_k$ at 
arbitrary occupation $\vec{q}$ of the array:
\begin{equation}
V_{t,k}(\vec{q}) = \frac{e}{2}\left(S^N-\frac{1}{C_k}\right)-
e\sum_{i=1}^{k-1}q_iS^i+e \sum_{i=k}^{N-1}q_iS_i.
\label{eq:evtk}
\end{equation}
The threshold voltage is determined as follows. For an initial charge state,
we determine the minimal activation energy $eV_{t,k}(\vec{q})$ to allow a 
tunneling event in the array, as well as the corresponding final charge state. 
The final charge state becomes the initial
state in the next step. The minimal activation energy for the new charge state
and the corresponding final charge state are again determined, and  
this procedure is repeated until one electron has been transported 
from emitter to collector. The largest of the activation energies found 
equals $eV_t$. 
In the special case that all background charges are zero, one has 
\begin{equation}
\label{eq:inhomC}
V_t=\case{1}{2}e\left(\sum_{i=1}^N 1/C_i-\mbox{Max}[1/C_1,1/C_2,\ldots,1/C_N]
\right), 
\end{equation}
which is an extension of the result $V_t=\case{1}{2}e\mbox{Min}[1/C_1,1/C_2]$ 
for a double junction.\cite{Amman} 
For $N\to \infty$, $V_t$ has a Gaussian distribution
with average $\case{1}{2}Ne\langle C^{-1}\rangle$ and variance $\mbox{Var}
V_t = \case{1}{4}Ne^2\mbox{Var} C^{-1}$.

\subsection{Turnstile configuration}
We next consider a turnstile configuration, i.e. an array with a single gate 
electrode coupled capacitively 
(capacitance $C_g$) to island $n$. The elements of the inverse capacitance 
matrix are then given by
\begin{eqnarray}
R_{i,j} & = & (S^i+C_g S^n S_n^i)S_j(S^N+C_g S^nS_n)^{-1}, 
\quad n\le i\le j\nonumber\\
R_{i,j} & = & S^i S_j(S^N+C_g S^nS_n)^{-1},\quad i\le n \le j\nonumber\\
R_{i,j} & = & S^i (S_j+C_g S_j^n S_n)(S^N+C_g S^nS_n)^{-1},
\quad i\le j\le n\nonumber\\
R_{j,i} & = & R_{i,j}.\label{eq:invcapmat}
\end{eqnarray}
In order to determine the threshold voltage $V_{t,k}(\vec{q})$, we have to 
distinguish between $k \le n$ and $k > n$. 
From Eqs.\ (\ref{eq:dG}) and (\ref{eq:invcapmat}) we find that 
$V_{t,k}(\vec{q})$ now depends on the gate voltage $V_g$:\cite{Nakazato}
\bleq
\begin{equation}
\label{eq:evtkgate}
V_{t,k}(\vec{q}) = \frac{e}{2}\left(S'^N-\frac{1}{C'_k}\right)
-e\sum_{i=1}^{k-1}q_iS'^i+e\sum_{i=k}^{N-1}q_iS'_i
+C_g[V_g-\case{1}{2}(V_e+V_c)]\times\left\{\begin{array}{ll}
S_n(1+\case{1}{2}C_gS_n)^{-1}, & k\leq n\\ 
-S^n(1+\case{1}{2}C_gS^n)^{-1}, & k>n\end{array}\right.
\end{equation}
\eleq
\noindent
where $S'$ is defined as in Eq.\ (\ref{eq:invsum}) in terms of modified 
capacitances $C'$:
\begin{equation}
\begin{array}{lr}
\vphantom{\frac{q}{q}}C_l' = C_l(1+\case{1}{2}C_g S_n)(1+C_g S_n)^{-1}, 
& k \le n, l \le n,\\
\vphantom{\frac{q}{q}}C_l' = C_l(1+\case{1}{2}C_g S_n), 
&  k\le n, l > n,\\
\vphantom{\frac{q}{q}}C_l' = C_l(1+\case{1}{2}C_g S^n), 
&  k > n, l \le n,\\
\vphantom{\frac{q}{q}}C_l' = C_l(1+\case{1}{2}C_g S^n)(1+C_g S^n)^{-1}, 
& k > n, l> n.
\end{array}
\end{equation}

\section{Background charge}
\label{sec:four}

The background charge in a single-electron tunneling device has a
large influence on its properties. For example, by tuning the background 
charge in a double junction with one gate one can set the threshold voltage
to any value between zero and $e/(2C+C_g)$.
In this section, we investigate the effect of background charges on the 
threshold voltage of an array of tunnel junctions.
For reasons of clarity, we choose identical junction capacitances in the
following ($C_i=C$ for all $i$). We start by investigating an array with a
non-zero background charge on a single island. We then give ensemble-averaged 
results for random background charges on all islands and compare with the 
experiments of Delsing et al.\cite{Delsing}

In the absence of gate coupling ($C_{g,i}=0$ for all $i$) and for a 
non-zero background charge $q_{0,m}=Q_{0,m}/e$ on island $m$, there are 
three initial tunneling events which may form the bottleneck for conduction: 
\begin{itemize}
\item{transfer of an electron from the emitter to the first island
(electron injection through junction $k=1$);}
\item{tunneling through junction $k=m+1$ if $q_{0,m} > 0$ or through junction 
$k=m$ if $q_{0,m} < 0$ (electron-hole creation at island $m$);}
\item{transfer from the last island to the collector (hole injection 
through junction $k=N$).}
\end{itemize}
An analysis of the corresponding tunneling sequences results in the 
threshold voltage
\begin{mathletters}
\begin{eqnarray}
V_t & = & \frac{e}{2C}\left(N-1-2\mbox{Min}[m q_{0,m}, 
(N-m)(1-q_{0,m})]\right),\nonumber\\
& & q_{0,m} \geq 0,\\
V_t & = & \frac{e}{2C}\left(N-1-2\mbox{Min}[m(1-|q_{0,m}|),(N-m)|q_{0,m}|]
\right),
\nonumber\\
& & q_{0,m} < 0.
\end{eqnarray}
\end{mathletters}%
For a uniform distribution of $q_{0,m}$ between $\pm \case{1}{2}$ and a 
uniform distribution of $m$ between 1 and $N-1$ its expectation value 
is $\langle V_t \rangle = (5N-7)e/12C$, with variance 
$\mbox{Var} V_t = (e/2C)^2 (N+1)(3N^2-5N+8)/180N$. The expectation
value is slightly smaller than for a homogeneous array without background
charges: $V_t = (N-1)e/2C$. 
In the limit $N\to\infty$ the root-mean-square deviation is $\mbox{rms}V_t 
\propto Ne/C$, of the same order as the threshold voltage itself.

\begin{figure}[htb]
\epsfxsize=0.95\hsize
\vspace*{-6ex}\epsffile{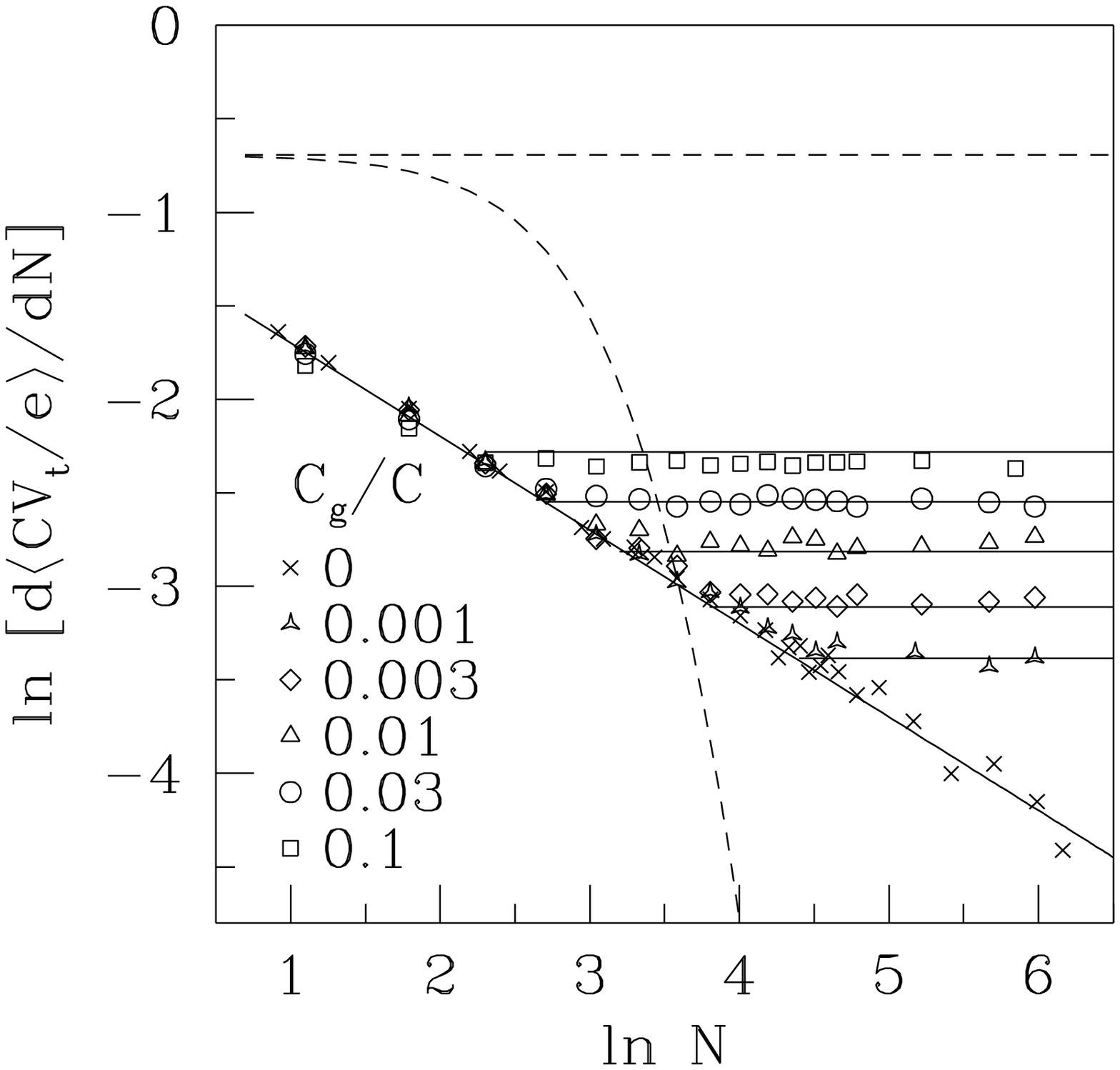}\vspace*{-4ex}
\refstepcounter{figure}
\label{fig:crossover}
{\small FIG. \ref{fig:crossover}.
Derivative of the average threshold voltage with respect to the
array length $N$, for ensembles of arrays with identical 
capacitances ($C_i=C$ and $C_{g,i}=C_g$ for all $i$) and random background 
charges on all islands, calculated from Eq.\ (\protect\ref{eq:allgates}). 
The average is determined numerically from ensembles of 10000 samples for 
$N\leq 128$ and ensembles of 1000 samples for larger arrays.
A cross-over from $\langle V_t\rangle \propto N^{1/2}$ to $\langle V_t\rangle
\propto N$ occurs at $N_c \approx 2.5\sqrt{C/C_g}$. Solid lines are the
extrapolation formulas (\protect\ref{eq:numvtsmallN}) and 
(\protect\ref{eq:numvtlargeN}). The dashed curves are obtained from the 
result (\protect\ref{eq:noq0}) for zero background charges and $C_g/C=0$ 
(upper curve) and $C_g/C=0.01$ (lower curve).}
\end{figure}
We next consider a one-dimensional array of equally
gated islands ($C_i=C$, $C_{g,i}=C_g$ for all $i$). In Refs.\ 
\onlinecite{Bakhvalov} and \onlinecite{Hu} the
charge transport in homogeneous arrays by soliton-like excitations was 
introduced. In terms of the soliton width 
$\lambda^{-1}=[2\mbox{arsinh}\sqrt{C_g/4C}]^{-1}$ of Ref.\ 
\onlinecite{Bakhvalov}, the threshold voltage for an electron tunneling 
through junction $k$ is given by
\bleq
\begin{eqnarray}
V_{t,k}(\vec{q}) & = & \frac{e}{2C} 
\left( -2\sum_{i=1}^{k-1}(q_i+q_g)
\sinh(i\lambda) \cosh[(N-k+\case{1}{2})\lambda]
+2\sum_{i=k}^{N-1}(q_i+q_g) \sinh[(N-i)\lambda]
\cosh[(k-\case{1}{2})\lambda]\right.\nonumber\\
&&\hphantom{\frac{e}{2C}(}\left.+\sinh[(N-\case{1}{2})\lambda]-\cosh[(N-2k+1)\lambda]
\sinh\frac{\lambda}{2}\vphantom{\sum_k^N}\right)
\left(\sinh\lambda\cosh\frac{N\lambda}{2} 
\cosh\frac{(N-2k+1)\lambda}{2}\right)^{-1}.
\label{eq:allgates}
\end{eqnarray}
\eleq%
\noindent
Here, the gate-induced charge $q_g\equiv C_g[V_g-\case{1}{2}(V_e+V_c)]$ acts
as an offset on the background charge. The average threshold voltage 
(averaged over the background charge) is 
therefore independent of $V_g$. For $N=2$, we find
\begin{equation}
\label{eq:n2av}
\langle V_t \rangle=e/(4C+2C_g).
\end{equation}

In the absence of background charges and for $q_g=0$, we find
\begin{equation}
\label{eq:noq0}
V_t=\frac{e}{2C} \frac{\sinh[(N-1)\lambda/2]}{\cosh(N\lambda/2) 
\sinh(\lambda/2)},
\end{equation}
which approaches a constant value as $N\to \infty$, provided $\lambda \not= 0$,
i.e. provided $C_g/C \not= 0$.
In Figure \ref{fig:crossover} we show the effect of random
background charges on all islands in arrays of different lengths for 
several gate couplings, as calculated from Eq.\ (\ref{eq:allgates}). 
The averages are computed numerically by putting a random charge $q_{0,k} \in 
(-\case{1}{2},\case{1}{2})$ on each island $k$. The dependence of 
$\langle V_t\rangle$ on the array length differs drastically from the 
result (\ref{eq:noq0}) without background charges: Instead of a threshold 
voltage which exponentially approaches a constant value as $N\to\infty$, we 
find $\langle V_t\rangle \propto \sqrt{N}-1$ for small arrays, with a 
cross-over to a linear $N$-dependence for large arrays. 
For $C_g\ll C$, the array length $N_c$ at which 
the cross-over occurs is found to be 2.5 times the soliton width,
\begin{equation}
N_c \approx 2.5\sqrt{C/C_g} \approx 2.5\lambda^{-1}.
\end{equation}

For $N < N_c$ the average threshold voltage is well described by an 
extrapolation of the result (\ref{eq:n2av}) for $N=2$:
\begin{equation}
\label{eq:numvtsmallN}
\langle V_t^<\rangle =\frac{e}{4C+2C_g}\frac{\sqrt{N}-1}{\sqrt{2}-1}.
\end{equation}
For $N>N_c$ we can describe the numerical data by
\begin{eqnarray}
\langle V_t^>\rangle & = & \langle V_t^<\rangle_{N=N_c}+ (N-N_c)\left.
\frac{\mbox{d} \langle V_t^<\rangle}{\mbox{d}N}\right|_{N=N_c}\nonumber\\
\label{eq:numvtlargeN}
& = & \frac{e}{4C+2C_g}\frac{1}{\sqrt{2}-1}
\left(\frac{N+N_c}{2\sqrt{N_c}}-1\right).
\end{eqnarray}
The cross-over to a linear $N$-dependence supports the intuitive idea that the 
background charge in the array is screened beyond $N_c$.
The rms deviation $\mbox{rms} V_t=0.31e (\sqrt{N}-1)/(2C+C_g)$ for all $N$.
The rms deviation of the threshold voltage for tunneling through a specific
junction $k$ has a much stronger dependence on $N$ than $\mbox{rms}V_t$ itself: 
$\mbox{rms}V_{t,k} \propto N^{3/2}$. Since $V_t$ is chosen as the 
maximal threshold voltage in a sequence of $N$ minimal values for single
tunneling events, the fluctuations in $V_t$ are smaller than those in $V_{t,k}$.
\begin{figure}[htb]
\epsfxsize=0.95\hsize
\vspace*{-10ex}\epsffile{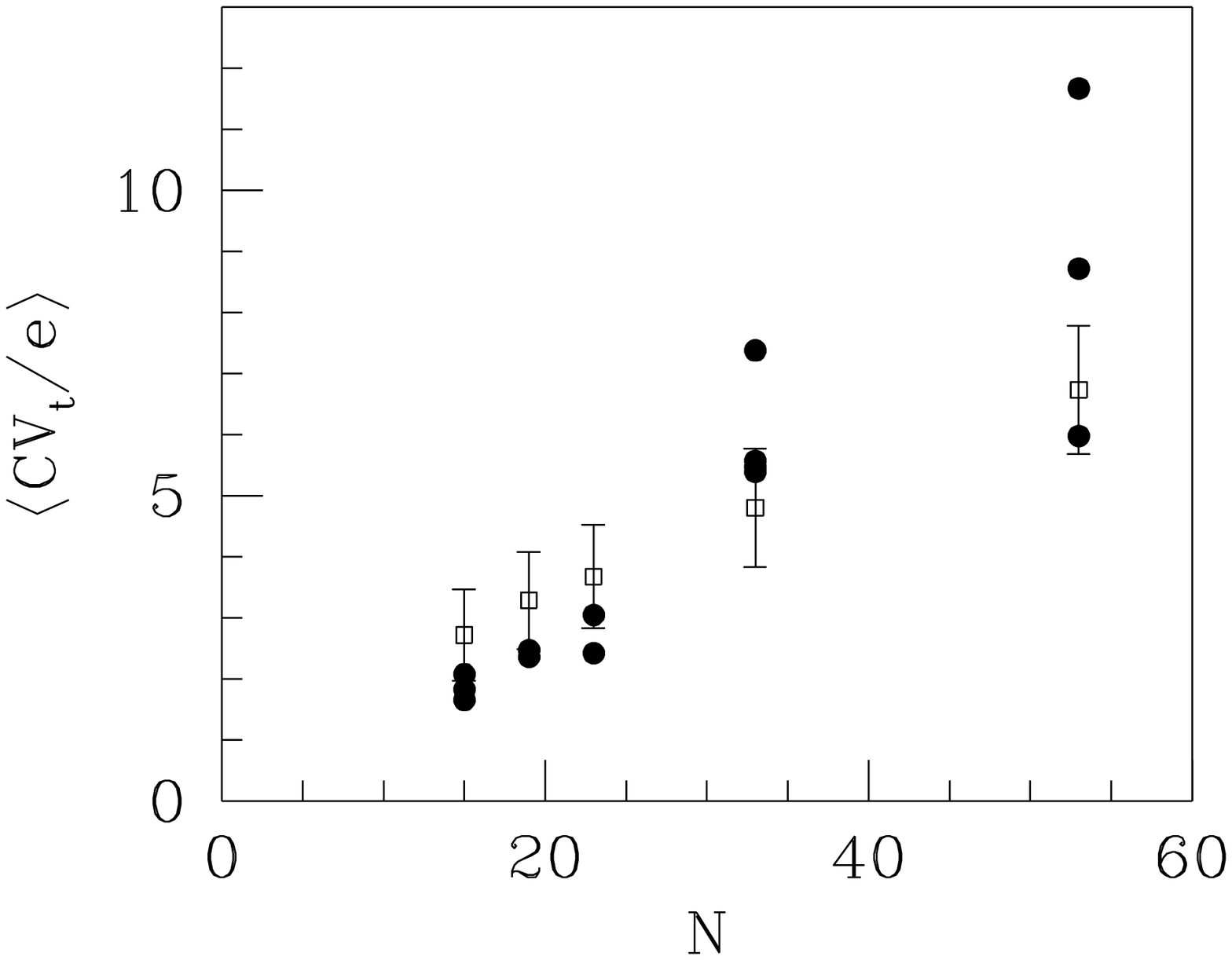}\vspace*{-4ex}
\refstepcounter{figure}
\label{fig:delsing}
{\small FIG. \ref{fig:delsing}.
Comparison of experimental threshold voltages (taken from Ref.\ 
\protect\onlinecite{Delsing}, solid dots) with the result of Eq.\ 
(\protect\ref{eq:allgates}), averaged over the random background charge 
(open squares with error bars). We used identical gate and junction 
capacitances, with $C_g/C=0.044$ ($N_c=12$), as estimated in Ref.\ 
\protect\onlinecite{Delsing}. There are no adjustable parameters.}
\end{figure}
\noindent
In Figure \ref{fig:delsing} we compare the threshold voltage
from Eq.\ (\ref{eq:allgates}), averaged over all background charges, with
experimental threshold voltages for arrays of different 
lengths.\cite{Delsing} We used the values $C=0.28$ fF and $C_g$=0.012 fF 
from Ref.\ \onlinecite{Delsing}, giving $N_c = 12$. Thus, the
experimental results are in the regime of a linear dependence of 
$\langle V_t\rangle$ on $N$.
The qualitative agreement is satisfactory, without any adjustable parameters.

In conclusion, we have derived an exact analytical expression for the 
threshold voltage $V_{t,k}(\vec{q})$ for tunneling through a junction $k$ in a 
one-dimensional array of $N$ metallic islands at arbitrary occupation 
$\vec{q}$ of the islands. We have calculated the average threshold voltage for 
transport and its fluctuations in a few simple cases. In particular, we have
found that including random background charges results in a $N^a$ dependence
of $\langle V_t\rangle$, with $a=\case{1}{2}$ for $N \stackrel{<}{~} 
2.5\sqrt{C/C_g}$ and $a=1$ for $N\stackrel{>}{~} 2.5\sqrt{C/C_g}$. We have 
made a comparison with the available experimental data on gated one-dimensional 
arrays,\cite{Delsing} and found a reasonable agreement. 

\begin{center}{\small \bf ACKNOWLEDGMENTS}\end{center}
Useful discussions with C. W. J. Beenakker are
gratefully acknowledged. A computer code to check our analytical results was
kindly made available to us by A. N. Korotkov. This work was supported by 
the Dutch Science Foundation NWO and by the Nordic Academy of Research
NORFA.

\ecols 

\end{document}